\documentclass[runningheads]{llncs}
\usepackage[T1]{fontenc}
\usepackage{graphicx}
\begin{document}
\title{Comparative Analysis of Community Detection Algorithms on the SNAP Social Circles Dataset
}
%
%
\author{Yash Malode\inst{1,4}\orcidID{0009-0006-9514-2483} \and
Amit Aylani \inst{1,5}\orcidID{0000-0002-9767-492X} \and Arvind Bhardwaj\inst{2,6}\orcidID{0009-0005-9682-6855} \and Deepak Hajoary \inst{3,7} 
}
%
%
\institute{Vidyalankar Institute of Technology, Mumbai, India. \and IEEE Senior Member \and Bodoland University, Assam, India. \and yash23malode@gmail.com \and prof.amitaylani@gmail.com \and arvind.bhardwaj@ieee.org \and hajoary.deepak@gmail.com
}
\maketitle              
\begin{abstract}

In network research, Community Detection has always been a topic of significant interest in network science, with numerous papers and algorithms proposing to uncover the underlying structures within networks. In this paper, we conduct a comparative analysis of several prominent community detection algorithms applied to the SNAP Social Circles Dataset, derived from the Facebook Social Media network. The algorithms implemented include Louvain, Girvan-Newman, Spectral Clustering, K-Means Clustering, etc. We evaluate the performance of these algorithms based on various metrics such as modularity, normalized cut-ratio, silhouette score, compactness, and separability. Our findings reveal insights into the effectiveness of each algorithm in detecting various meaningful communities within the social network, shedding light on their strength and limitations. This research contributes to the understanding of community detection methods and provides valuable guidance for their application in analyzing real-world social networks.

\end{abstract}
\section{Introduction}

In recent years, the rise of social media platforms like Facebook, Twitter, Instagram, and LinkedIn has fundamentally reshaped the dynamics of human interaction and information exchange. These platforms have become essential tools for global connectivity, enabling individuals and entities to engage, share, and disseminate information on an unprecedented scale. 
The proliferation of social media platforms and the volume of data they create from user interactions have given rise to a growing body of study devoted to the analysis of the complex networks and relationships that underlie these platforms. Understanding the structural patterns and behaviors within social networks holds significant promise for applications ranging from targeted marketing to the identification of influential communities and individuals.

The task of community detection, which entails locating coherent user groups with shared interests or strong connections, is essential to the research of social media networks. Community detection algorithms play a critical role in uncovering these underlying structures, but the sheer diversity of available algorithms poses challenges in selecting the most suitable approach for a given dataset. With numerous algorithms employing distinct methodologies and assumptions, determining the optimal algorithm for a specific social network dataset remains a complex endeavor.

In our research, we conduct a comparative analysis of prominent community detection algorithms using the SNAP Social Circles dataset, a subset of the Facebook social media network. This dataset comprises of anonymized ego networks representing friendships among Facebook users. Our study aims to assess the performance of these algorithms across various metrics such as modularity, normalized cut ratio, silhouette score, compactness, and separability. By systematically evaluating the efficacy of different algorithms in identifying communities within the social network, our research aims to offer insights into their respective strengths, weaknesses, and applicability in real-world scenarios. Through this endeavor, we seek to advance the understanding of community detection methods and provide valuable guidance to researchers, practitioners, and organizations seeking to leverage social media data for diverse purposes, including marketing, social network analysis, and beyond.

\section{Literature Review}
Modern-day network research has made great strides towards understanding complex systems, and community structure has become an important tool for analyzing graphs that model real-world systems. The organizing of vertices into clusters with sparse connections between clusters and dense interior connections reflects the compartmentalization seen in many systems, such as biological organisms' tissues or organs. Detecting communities holds substantial importance across disciplines such as sociology, biology, and computer science, where networks serve as fundamental representations of systems.

Fortunato's paper "Community Detection in Graphs" \cite{cd_in_graphs_Fortunato} offers a thorough summary of the challenges involved in community detection, emphasizing the need to delineate key elements, explore methodologies, and address critical issues such as clustering significance and method evaluation. With graph clustering lacking precise definitions, the literature presents a plethora of algorithms and approaches, spanning graph partitioning, modularity-based methods, spectral algorithms, dynamic techniques, and more. Modularity-based methods like Newman-Girvan modularity and spectral algorithms have garnered attention for their efficiency, while dynamic algorithms and statistical inference methods offer promising avenues. Despite progress, challenges persist in algorithm evaluation, understanding clustering in real-world contexts, and exploring dynamic community structures. The evolving field continues to refine existing methods, explore new approaches, and illuminate the properties and applications of communities in real-world networks.

"Community detection in networks: A user guide," \cite{cd_in_networks_Fortunato} offers an in-depth exploration of community detection tailored for practitioners and accessible to those with basic network science knowledge. Rather than striving for exhaustive coverage, the paper focuses on elucidating fundamental aspects of community detection, organized into three main sections. The first section explores conceptual foundations, tracing the evolution of the community concept and laying the groundwork for understanding diverse interpretations. The second section tackles validation challenges, emphasizing the importance of validation and discussing various validation methods. The final section delves into algorithmic approaches, critically examining popular clustering algorithms and methodological considerations. Additionally, the paper provides guidance on accessing software tools and concludes with key findings and suggestions for future research. Overall, it serves as a valuable resource for navigating community detection complexities, addressing both theoretical foundations and practical considerations.

The paper "Community Detection in Social Networks" by Bedi et al. \cite{cd_in_social_networks_Bedi} underscores the importance of community detection within the expansive realm of social networking. As individuals increasingly engage in virtual communities within social networking sites, understanding and identifying these clusters become paramount for various purposes, from collaborative research to targeted marketing strategies. Through a comprehensive survey of existing algorithms, categorized by their methodologies and applications across domains, the paper navigates through the fundamental concepts of social networks and community structure. By highlighting the potential applications and inclusivity of datasets utilized by these algorithms, the authors emphasize the versatility and significance of community detection, offering insights into its effective utilization across commercial, educational, and developmental landscapes.

The paper "Community detection algorithms: A comparative analysis" by Lancichinetti and Fortunato \cite{cda_comp_analysis_Fortunato} addresses the complexity of community detection in networks and the necessity for rigorous evaluation of algorithmic performance. Recognizing the limitations of previous evaluations, which often relied on small or simplified networks, the authors conduct comprehensive tests using diverse benchmarks to assess algorithmic accuracy and computational efficiency. Among the tested algorithms, Rosvall and Bergstrom's Infomap method emerges as particularly effective, especially on challenging benchmarks. However, the paper acknowledges ongoing challenges, such as the need for algorithms capable of handling hierarchical structures and multipartite graphs, highlighting avenues for further research in the field.

\section{Problem Statement}
Community Detection in social networks is a fundamental task crucial for understanding network structure and behavior. As social networks continue to grow exponentially, the need for efficient and accurate community detection algorithms becomes increasingly critical. Our aim in this research paper is to conduct a comprehensive comparative analysis of community detection algorithms using the SNAP Social Circles Dataset. By evaluating various algorithms' performance metrics such as modularity, normalized cut ratio, and silhouette score, we seek to identify the most effective algorithms for detecting communities within this dataset. This comparative analysis aims to provide insights into the strengths and weaknesses of different algorithms, aiding researchers and practitioners in selecting appropriate methods for analyzing large-scale social network data. Through this research, we aim to contribute to advancing community detection techniques and facilitate deeper insights into the structure and dynamics of social networks.

\section{Methodology}

\subsection{Terminologies}

\subsubsection{Network}:
A network, also known as a graph, is a collection of nodes (vertices) and edges (links) that represent relationships or interactions between the nodes. Networks can be directed (edges have a specific direction) or undirected (edges have no specific direction), and they can vary in size and complexity, ranging from small-scale social networks to large-scale infrastructure networks.

\subsubsection{Node and Edges}:
In context of networks, a node (or vertex) represents an individual entity within the network. An edge (or link) in a network connects two nodes and represents a relationship or interaction between them.

\subsubsection{Community}:
A community (or cluster) refers to a subset of nodes that are densely connected but sparsely connected to nodes outside the subset. Communities represent cohesive groups of nodes that exhibit similar properties or behavior within the network. Community detection algorithms aim to identify these groups or clusters to uncover underlying structures or patterns within the network.

\subsubsection{Degree}:
The degree of a node in a graph represents the number of edges connected to that node. In directed graphs, nodes have both an in-degree (number of incoming edges) and an out-degree (number of outgoing edges).

\subsubsection{Centrality}:
Centrality measures the importance or influence of a node within a network. Different centrality measures exist, including degree centrality (based on the number of connections), betweenness centrality (based on the shortest paths that pass through the node), closeness centrality (based on the average distance to all other nodes), and eigenvector centrality (based on connections to high-scoring nodes).

\subsubsection{Hub}:
A hub refers to a node with a high degree of connection to other nodes. Hubs play a central role in network communication and information flow, often serving as critical points for connectivity within the network. 

\subsubsection{Path}:
A path in a graph is a collection of nodes connected by edges, where each node is adjacent to the next node in the sequence. Paths can be directed or undirected. They can vary in length from a single edge to a series of interconnected nodes.

\subsubsection{Cycle}:
In a graph, a cycle is a closed path that travels through a series of nodes and edges without repeating any nodes, with the starting and finishing nodes usually the same. Cycles are fundamental components of graphs and can have important implications for network understanding.

\subsubsection{Connected Components}:
In a graph, subsets of nodes that have a network of edges connecting each node to every other node are known as connected components. If there are isolated node groups that are not connected to one another, then a graph may contain more than one connected component.

\subsection{Metrics}

\subsubsection{Modularity}:
Modularity is a measure that quantifies the strength of the community structure in a network. It evaluates the density of connections within communities relative to connections between communities. Higher modularity values indicate a more pronounced community.

\begin{equation}
Q = \frac{1}{2m}\sum_{i,j}[A_{ij} - \frac{k_ik_j}{2m}]\delta(c_i,c_j)
\end{equation}

\subsubsection{Normalized Cut Ratio}:
The normalized cut ratio is a measure used to evaluate the quality of a partition of nodes into communities. It assesses the balance between intra-community connections and inter-community connections, aiming to minimize the number of edges cut between communities while maximizing the number od edges within communities. Lower normalized cut ratios indicate better community partitions with cohesive communities and minimal inter-community connections.

\begin{equation}
Normalized\ Cut\ ({\pi}) = \frac{1}{k}\sum_{i=1}^k\frac{cut(C_i, C^{'}_{i})}{vol(C_i)}
\end{equation}

\subsubsection{Silhouette Score}:
The silhouette score is a metric used to assess the quality of clustering in a dataset. It measures how similar an object is to its own cluster compared to other clusters. A silhouette score ranges from -1 to 1, where a score closer to 1 indicates that the object is well-matched to its own cluster and poorly matched to the neighboring cluster. A score of near 0 indicates overlapping clusters and a score of less than 0 indicates misclassification.

\begin{equation}
S = \frac{\sum_{i=1}^N\frac{b_i - a_i}{max(b_i, a_i)}}{N}
\end{equation}
where 
\begin{equation}
b_i = min_{k\neq i}\frac{1}{|C_k|}\sum_{j\subset C_k}d(i,j)
\end{equation}
\begin{equation}
a_i = \frac{1}{|C_i|-1}\sum_{j\subset C_i, i\neq j}d(i,j)
\end{equation}

\subsubsection{Compactness}:
Compactness refers to the degree of cohesion within a community. A highly compact community consists of nodes that are densely interconnected, with strong internal connections and minimal external connections. Lower values of compactness indicate tighter and more compact clusters, while higher values indicate greater dispersion or spread of data points within clusters.

\begin{equation}
Compactness\ = \sum_{i=1}^k\frac{1}{|C_i|}\sum_{x\subset C_i}dist(x,\mu_i)^2
\end{equation}

\subsubsection{Calinski-Harabasz Score}:
The Calinski-Harabasz score, also known as the variance ratio criterion, is a metric used to evaluate the quality of clustering in a dataset. It measures the ratio of between-cluster dispersion to within-cluster dispersion, aiming to maximize the separation between clusters while minimizing the dispersion within clusters. Higher Calinski-Harabasz scores indicate better-defined and more compact clusters.

\begin{equation}
CH\ = (\frac{B}{W})(\frac{N-k}{k-1})
\end{equation}
where

B = between-cluster dispersion

W = within-cluster dispersion

N = Total data points

k = Number of clusters

\subsubsection{Separability}
It measures the degree of distinction between communities. It evaluates the extent to which communities can be differentiated from one another, with well-separated communities having minimal overlap and clear boundaries.

\begin{equation}
Separability\ = \frac{1}{k(k-1)}\sum_{i=1}^k\sum_{j\neq i}dist(\mu_i, \mu_j)
\end{equation}

\section{Implementation}

\subsection{Dataset}

\subsubsection{SNAP Social Circles: Facebook}:

This dataset consists of 'circles' (or 'friends lists') from the Facebook social media site. The data was collected from survey participants using the Facebook app. The information has been made anonymous by substituting new values for each user's Facebook internal ID. The dataset consists of more than 4000 nodes having more than 88,000 edges.

\subsection{Workflow}

\begin{figure}
\centering
\includegraphics[width=300pt]{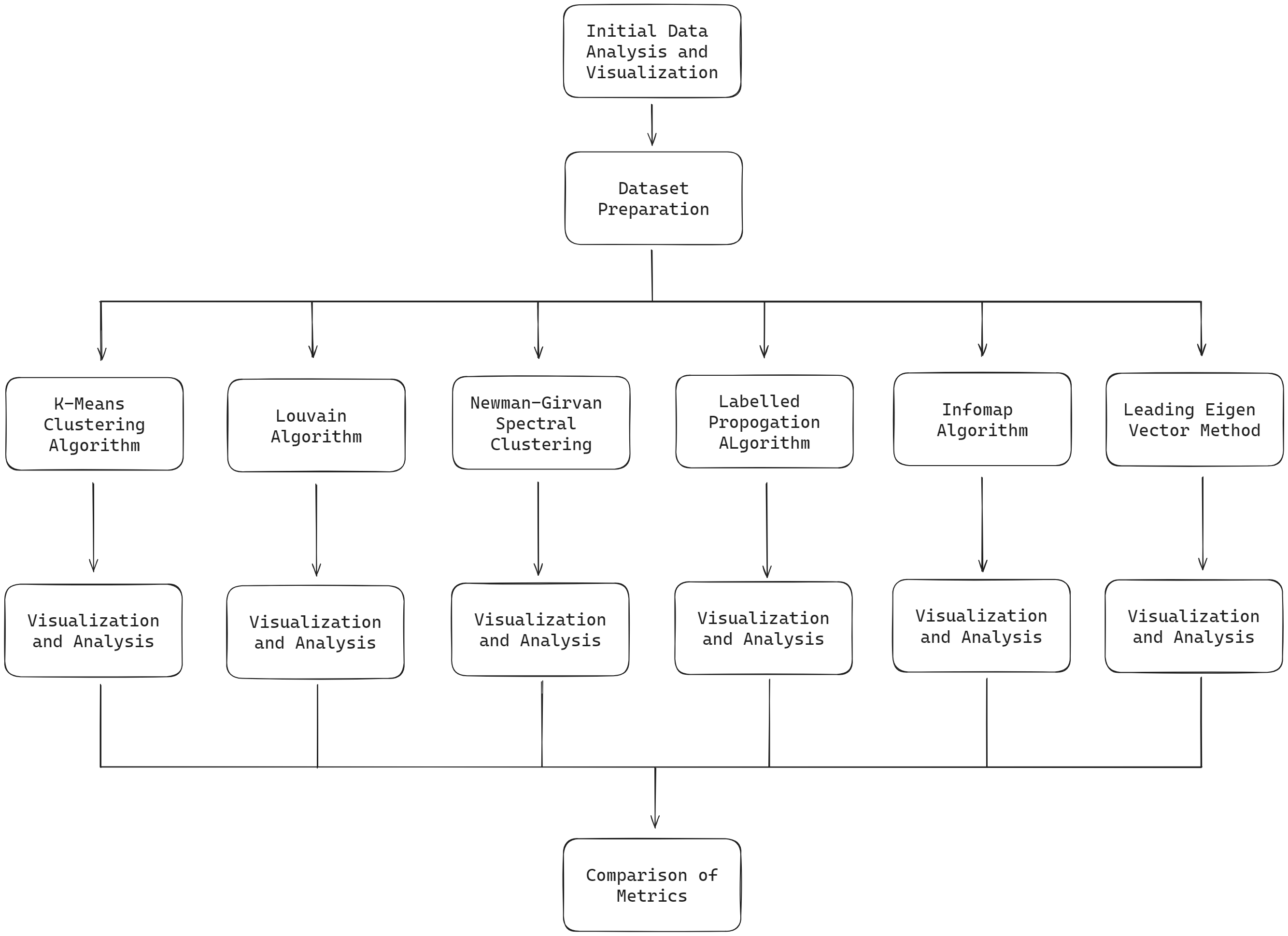}
\caption{Approach Used for Implementation}
\end{figure}

\subsection{Data Analysis}

A quick, initial analysis shows more than 95\% of nodes have a degree lying between 0 and 200. While a few nodes - possibly the major hub nodes have a much higher degree. (See Figure 2 and Figure 3)

\begin{figure}
\centering
\includegraphics[width=250pt]{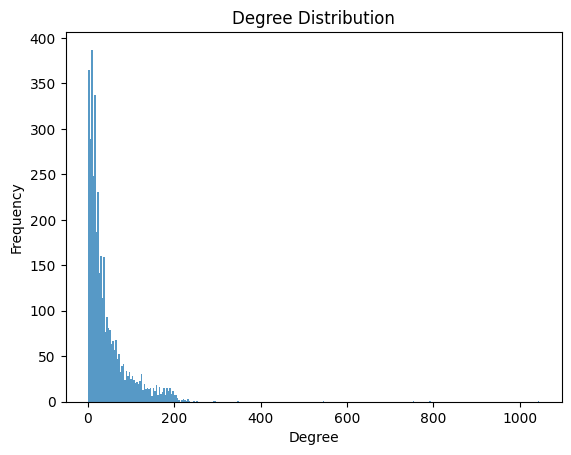}
\caption{Degree Distribution of nodes in the dataset}

\end{figure}

\begin{figure}
\centering
\includegraphics[width=250pt]{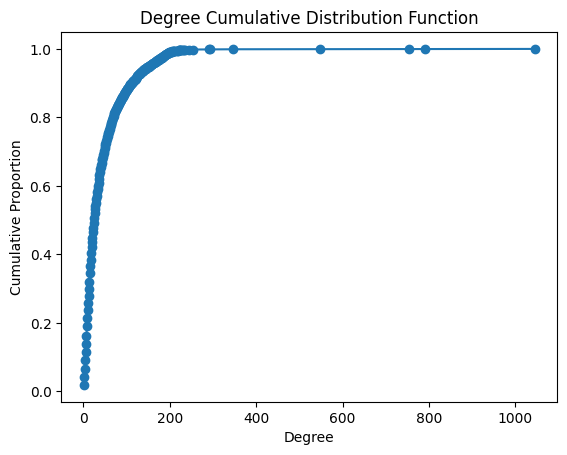}
\caption{Cumulative Distribution Plot of nodes in the dataset}
\end{figure}

To know more about these potential hubs, a scatterplot can be used to find nodes with the highest degree centrality. These hubs often act as central points of communication, facilitating flow of information between different parts of the network. In many real-world networks, hubs often emerge naturally due to many factors. 

However, it is important to note that although hubs are frequently nodes with high degree centrality, not all hubs are necessarily nodes with high degree centrality, and vice versa. Many other factors influence a node's importance in networks. (See Figure 4)

\begin{figure}
\centering
\includegraphics[width=200pt]{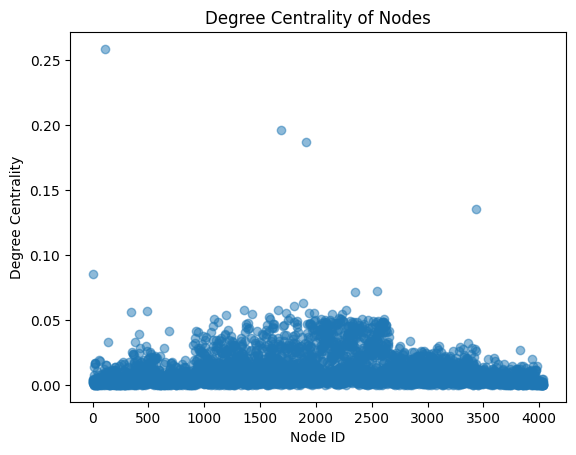}
\caption{Degree Centrality Plot of each node}
\end{figure}

\subsection{Initial Visualization}
In our research, we employed the matplotlib library to conduct preliminary visualizations of the SNAP dataset. Matplotlib, a widely-used Python plotting library, offers a versatile toolkit for creating a variety of plots and charts. Leveraging its functionality, we were able to generate insightful visualizations that provided initial insights into the structure and characteristics of the dataset.

By utilizing matplotlib for our visualization needs, we were able to lay the groundwork for further analysis and interpretation of the SNAP dataset, setting the stage for more in-depth explorations and insights in our research.

\begin{figure}
\centering
\includegraphics[width=300pt]{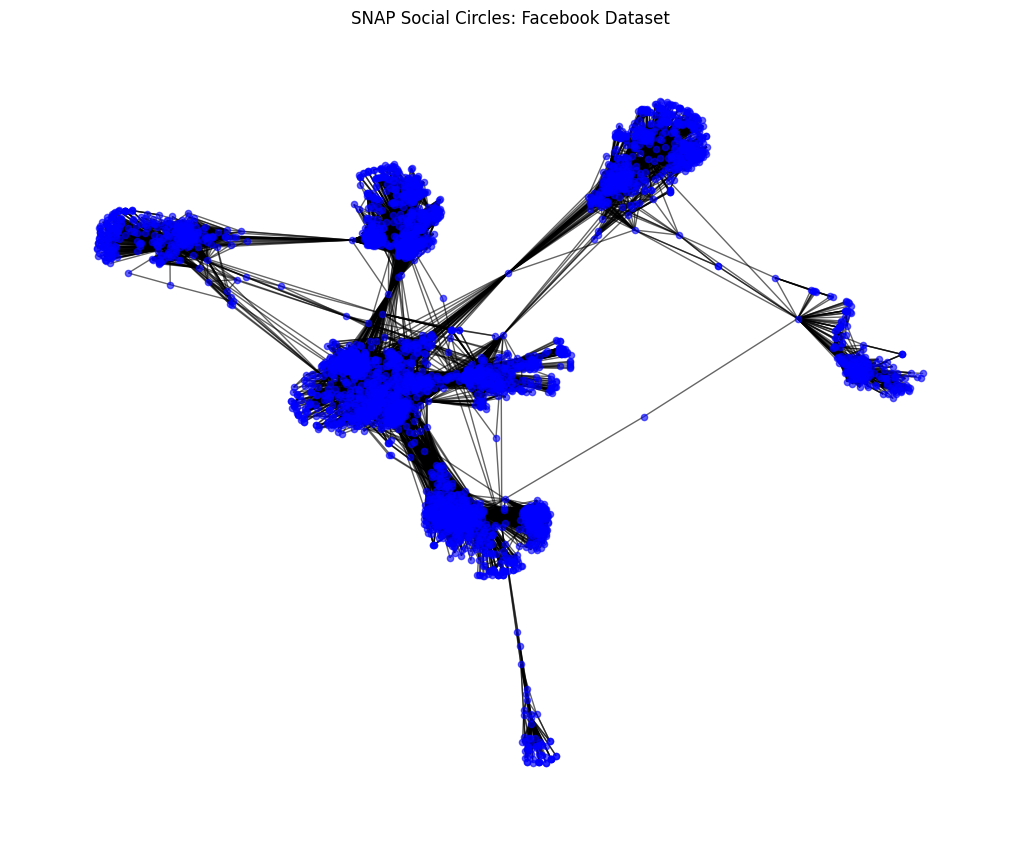}
\caption{}
\end{figure}

\subsection{Algorithms}

\subsubsection{A) K-Means Clustering Algorithm}:

\begin{equation}
arg\ min_S\sum_{i=1}^k\sum_{x\subset S_i}||x-\mu_i||^2\ =\ arg\ min_S\sum_{i=1}^k|S_i|Var\ S_i
\end{equation}
where
\begin{equation}
\mu_i = \frac{1}{|S_i|}\sum_{x\subset S_i}x
\end{equation}

K-means Clustering aims to partition n observations into k clusters in which each observation belongs to the cluster with the nearest mean, serving as a prototype of the cluster. \cite{kmeans} \cite{kmeans2020}

After experimenting with various values of K in the K-means clustering algorithm, it was determined that setting the number of clusters to 15 yielded the most optimal results. (See Figure 6)
\begin{figure}
\centering
\includegraphics[width=200pt]{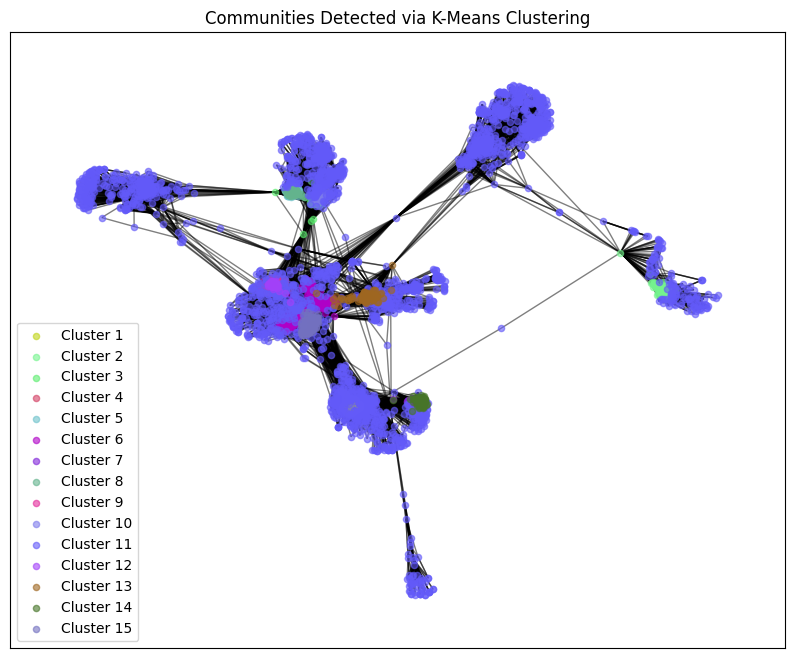}
\caption{Communities detected by KMeans Clustering}
\end{figure}

\subsubsection{B) Louvain Method for Community Detection}:

The Louvain method, developed by Vincent D. Blondel et al. \cite{louvain}, is a popular and efficient algorithm for community detection in complex networks. It is based on the optimization of modularity, a metric that measures the quality of a partition of a network into communities. The Louvain method employs a greedy optimization strategy, iteratively merging smaller communities into larger ones to maximize the overall modularity score. 
The scalability and capability of the Louvain method to manage massive networks with millions of nodes and edges is well-known due to its iterative approach, being able to identify communities of varying sizes efficiently. Its effectiveness and speed make it widely used in various fields, including social network analysis, biological network analysis, and more \cite{louvainafter15}. A total of 13 communities were detected by Louvain Algorithm which has been visualized below. (See Figure 7)
\begin{figure}
\centering
\includegraphics[width=200pt]{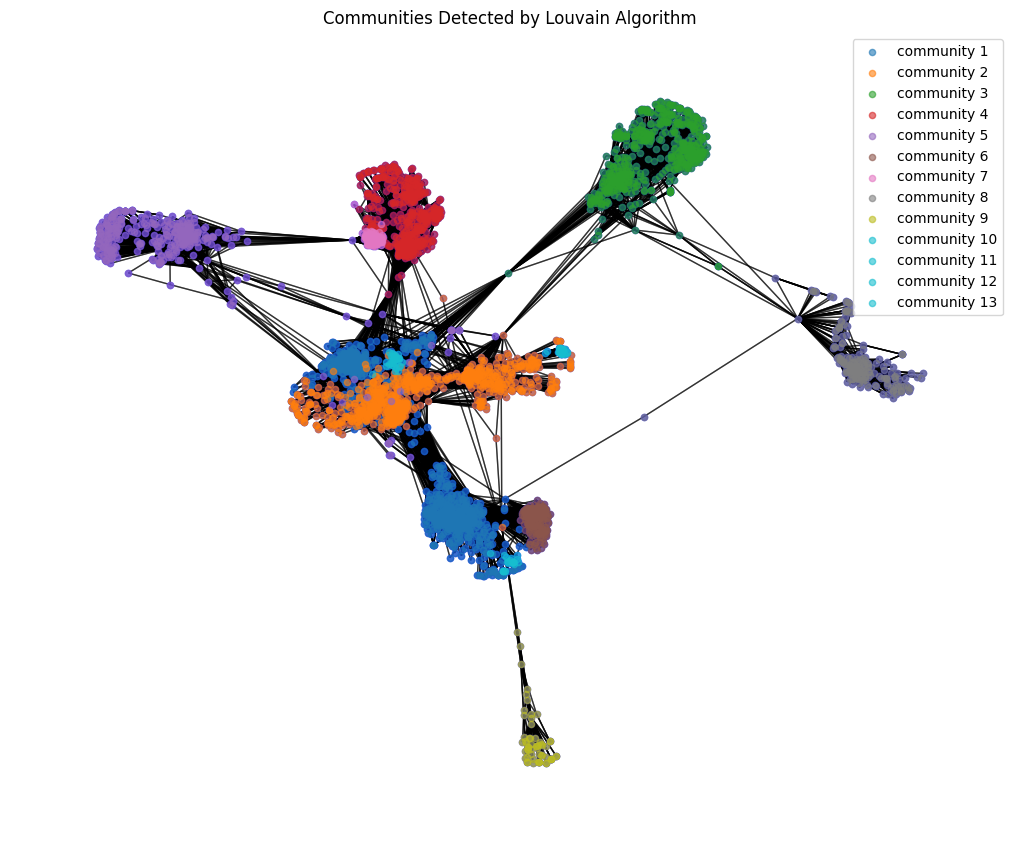}
\caption{Communities detected by Louvain Algorithm}
\end{figure}

\subsubsection{C) Spectral Clustering}:

Spectral clustering is an effective method frequently used for community detection in network analysis. It leverages the spectral properties of the network's adjacency matrix or Laplacian matrix to partition the nodes into distinct communities. By computing the eigenvectors corresponding to the largest eigenvalues of these matrices and embedding the nodes into a lower-dimensional space, spectral clustering aims to identify densely connected regions in the network. 

This method is particularly effective for identifying communities with complex structures, such as overlapping or hierarchical communities. Spectral clustering offers several advantages, including its ability to handle networks with irregular shapes and its flexibility in accommodating different types of connectivity patterns. Additionally, spectral clustering can be applied to both undirected and directed networks, making it a versatile tool for community detection across various domains, including social networks, biological networks, and more \cite{specclustering2020} \cite{onspecclustering} \cite{specclusteringtutorial}. The communities detected by Spectral clustering have been visualized below. (See Figure 8)
\begin{figure}
\centering
\includegraphics[width=275pt]{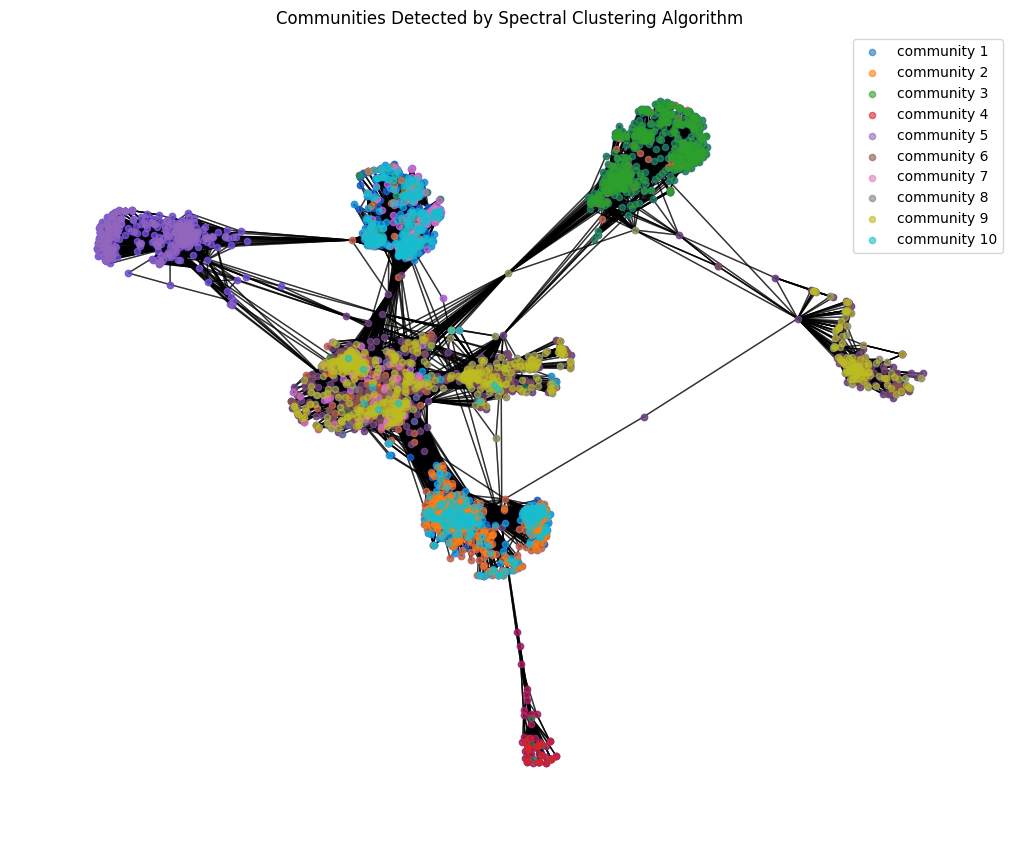}
\caption{Communities detected by Spectral Clustering}
\end{figure}

\subsubsection{D) Label Propagation Algorithm}:

Label propagation is a simple yet effective algorithm for community detection in networks. It operates by iteratively updating the community assignments of nodes according to the labeling of their neighbors. Each node is initially given a unique label, and in each iteration, nodes adopt the most common label among their neighbors. This process continues until convergence, where the labels stabilize and each node becomes associated with a specific community. Label propagation is known for its scalability and computational efficiency, making it suitable for large-scale networks.

However, it may struggle with certain network structures, such as networks with highly interconnected communities or networks with low-density regions. Despite its simplicity, label propagation has been widely used in various applications, including social network analysis, citation networks, and recommendation systems, where identifying communities is crucial for understanding network structure and dynamics \cite{labprop} \cite{labpropraghavan}.

The Label Propagation Algorithm identified 44 communities within the network, with some communities containing fewer than 10-15 nodes. To enhance visualization clarity, only the top 15 communities with the highest membership have been depicted in the plots. (See Figure 9)
\begin{figure}
\centering
\includegraphics[width=275pt]{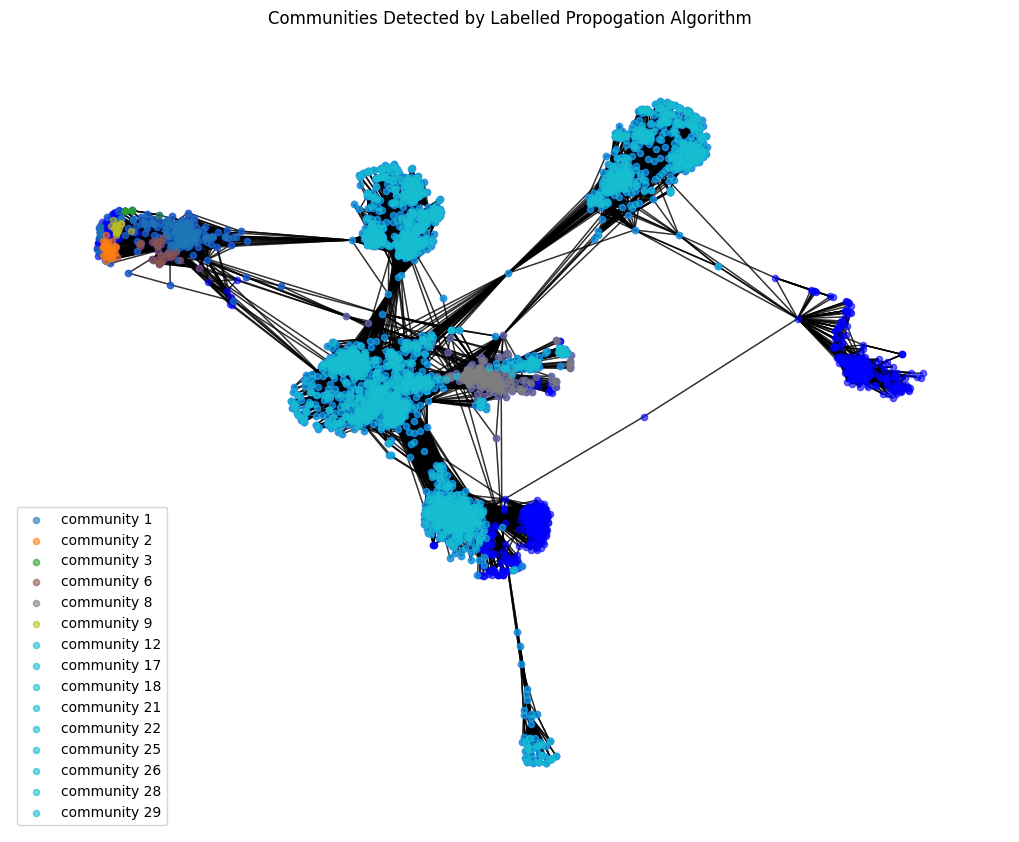}
\caption{Communities detected by Label Propagation Algorithm}
\end{figure}

\subsubsection{E) Infomap Algorithm}:

The Infomap algorithm is a widely used technique for detecting communities in networks, particularly in the realm of complex systems and network science. Developed by Martin Rosvall and Carl T. Bergstrom, Infomap is founded on the notion of optimizing a map of information flow within a network. It treats the network as a flow of random walkers, where nodes represent states and edges represent possible transitions between states. The algorithm aims to find the partitioning of the network that compresses the information flow representation the most, effectively identifying densely connected groups of nodes as communities. 

Infomap's strength lies in its ability to uncover hierarchical and overlapping community structures, providing insights into the multi-scale organization of complex networks. Furthermore, Infomap is known for its efficiency and scalability, making it applicable to large-scale networks commonly encountered in real-world systems such as social networks, biological networks, and information networks. Its widespread adoption and successful application across diverse domains underscore its importance in the field of network analysis and community detection \cite{infomapalgo}.

The Infomap Algorithm identified 93 communities within the network, with many communities containing fewer than 10 nodes. To enhance visualization clarity, only the top 15 communities with the highest membership have been depicted in the plots. (See Figure 10)
\begin{figure}
\centering
\includegraphics[width=275pt]{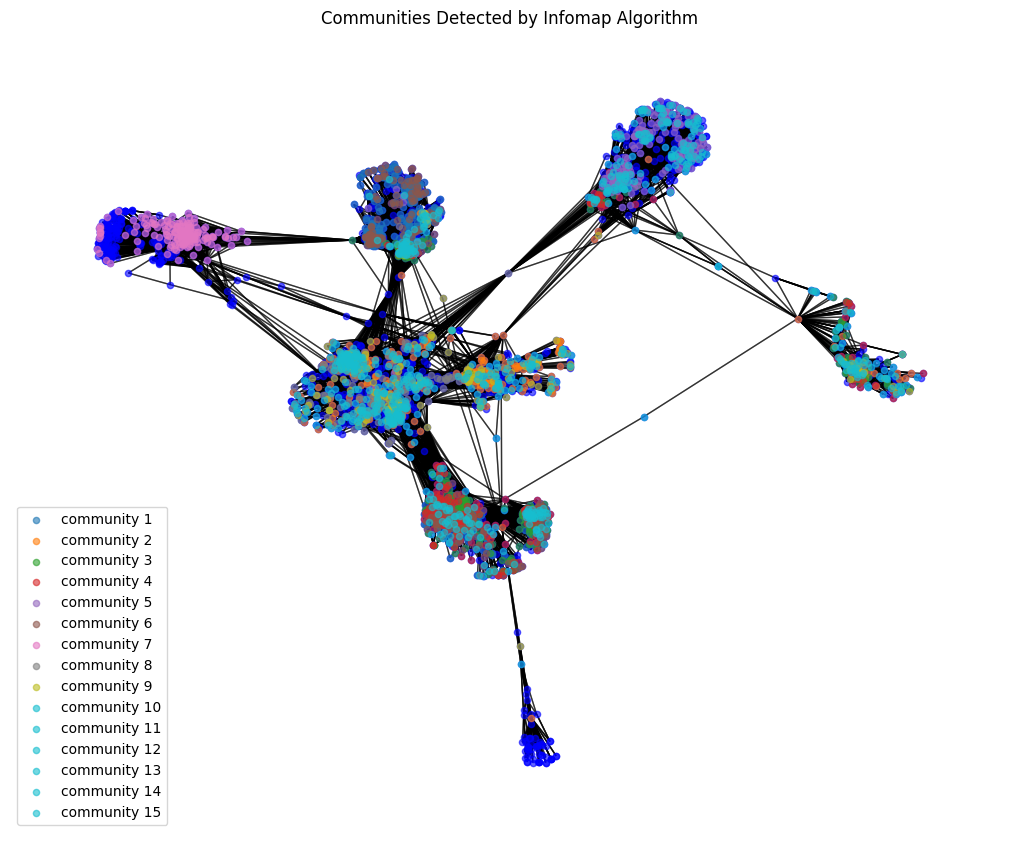}
\caption{Communities detected by Infomap Algorithm}
\end{figure}

\subsubsection{F) Leading Eigenvector Algorithm}:

The leading eigenvector algorithm, frequently called the Newman-Girvan algorithm, is a prominent method for community detection in complex networks. Developed by Mark Newman and Michelle Girvan, this algorithm leverages the spectral properties of a network's adjacency matrix to identify communities based on the eigenvector corresponding to the largest eigenvalue. 

The basic idea behind the algorithm is to iteratively partition the network by splitting it along the direction of the leading eigenvector of the adjacency matrix, which tends to capture the network's underlying community structure. By recursively applying this process, the algorithm identifies cohesive groups of nodes that are densely connected internally while sparsely connected to nodes in other communities. The leading eigenvector algorithm is known for its simplicity, efficiency, and effectiveness in detecting communities in various types of networks, including social networks, biological networks, and communication networks. 

Despite its straightforward approach, the algorithm has demonstrated remarkable performance and has become a foundational tool in the field of network analysis and community detection \cite{leadingevec}. The Infomap Algorithm identified 18 communities within the network, with some communities containing fewer than 10 nodes. To enhance visualization clarity, only the top 15 communities with the highest membership have been depicted in the plots. (See Figure 11)
\begin{figure}
\centering
\includegraphics[width=250pt]{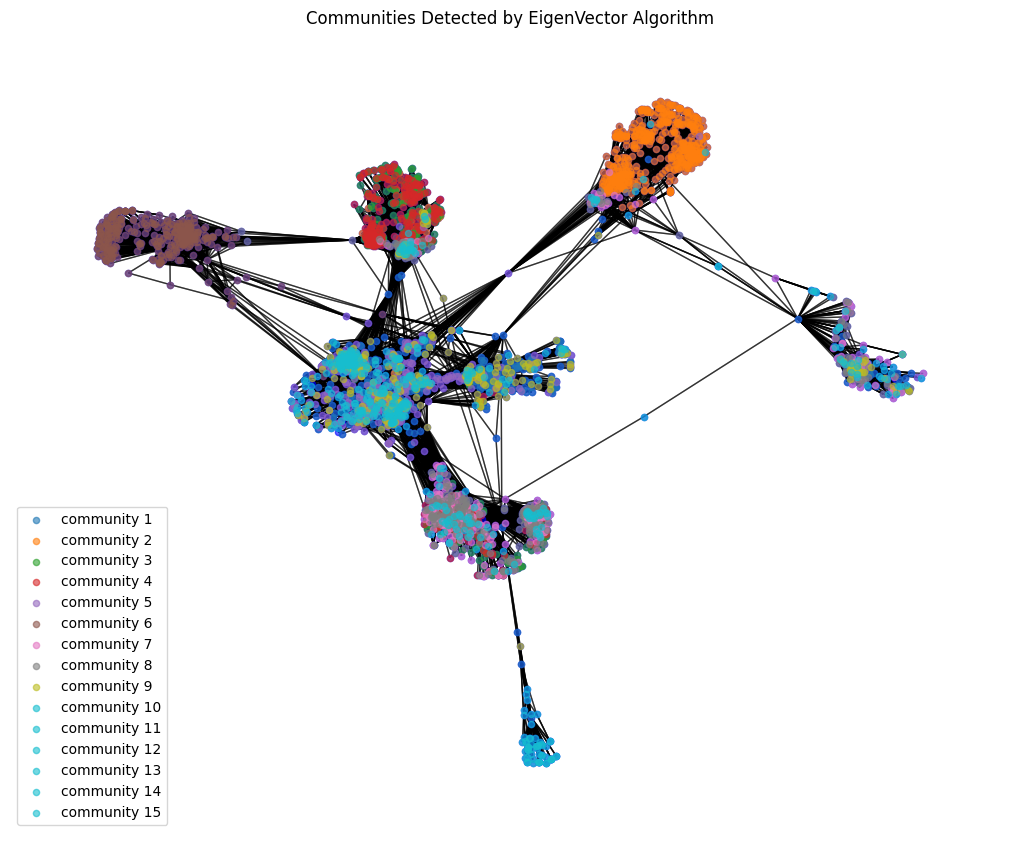}
\caption{Communities detected by Leading Eigen Vector Algorithm}
\end{figure}

\section{Comparison of Algorithms}

\subsection{Overview of all metrics}

The table below shows the metric values of all algorithms applied in our implementation. As mentioned above, six community detection algorithms have been tested on five separate metrics. (See Figure 12)

\begin{figure}
\centering
\includegraphics[width=400pt]{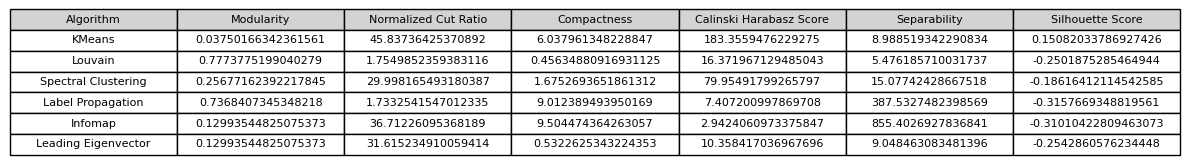}
\caption{An overview of all metric values}
\end{figure}

\subsection{Graphical Visualization of comparisons}

Here we compare all the algorithms with respect to each metric using radar charts. By plotting each algorithm as a separate axis and representing the metrics as spokes emanating from the center, we are able to present a comprehensive comparison of algorithm performance. The radar charts allow us to easily identify the strengths and weaknesses of each algorithm across different metrics, facilitating nuanced insights into their respective effectiveness in addressing specific objectives.
Since the value of the Silhouette Score ranges from -1 to 1, it has been visualized using a bar plot.(See Figure 13 to Figure 18)

\begin{figure}
\centering
\includegraphics[width=180pt]{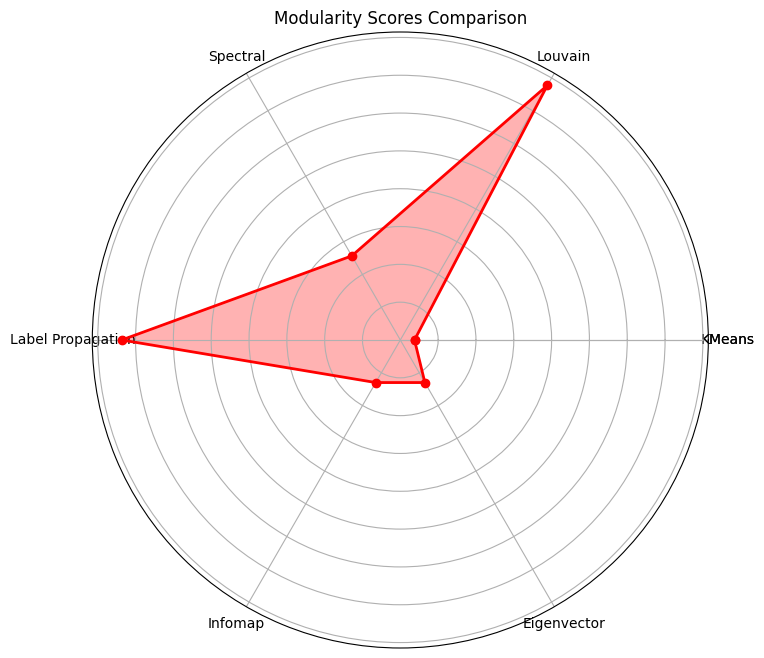}
\caption{Comparison of Modularity Score of the algorithms (Higher values are better)}
\end{figure}

\begin{figure}
\centering
\includegraphics[width=200pt]{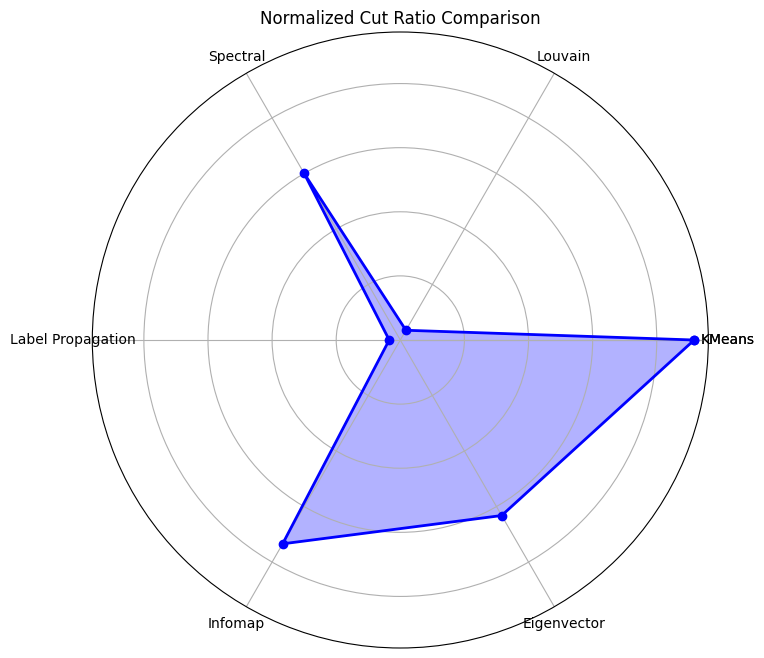}
\caption{Comparison of Normalized Cut Ratio of the algorithms (Higher values indicate more edges between communities)}
\end{figure}

\begin{figure}
\centering
\includegraphics[width=200pt]{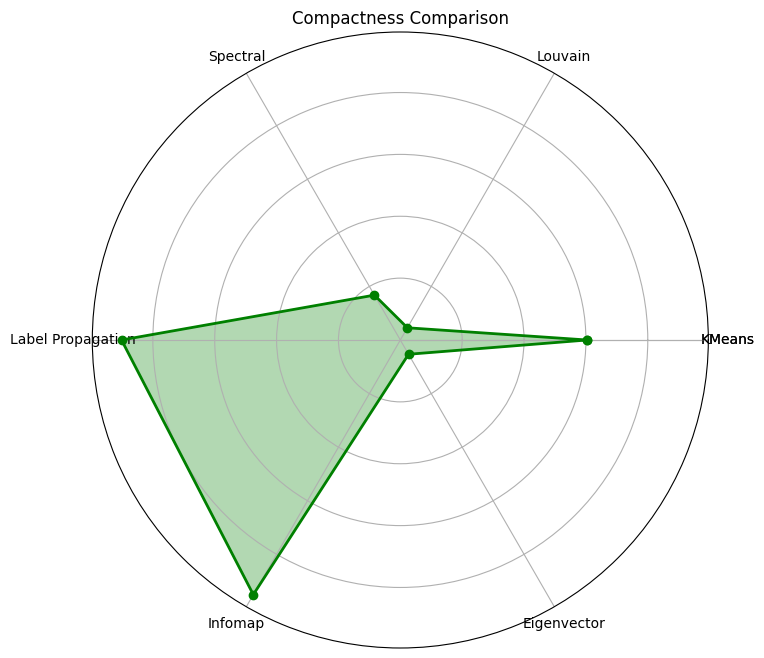}
\caption{Comparison of Compactness Score of the algorithms (Higher values indicate more cohesion within communities)}
\end{figure}

\begin{figure}
\centering
\includegraphics[width=200pt]{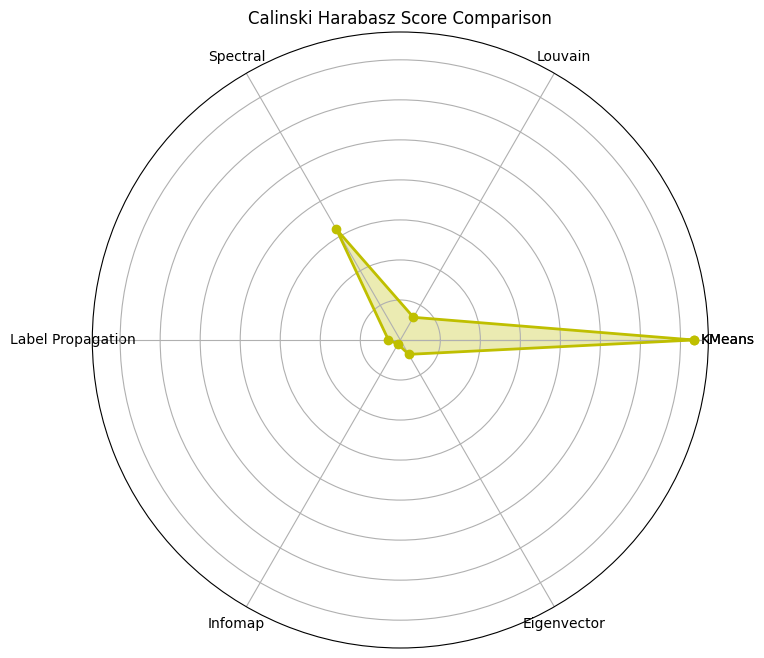}
\caption{Comparison of Calinski-Harabasz Score of the algorithms (Higher values indicate more defined, distinct communities)}
\end{figure}

\begin{figure}
\centering
\includegraphics[width=200pt]{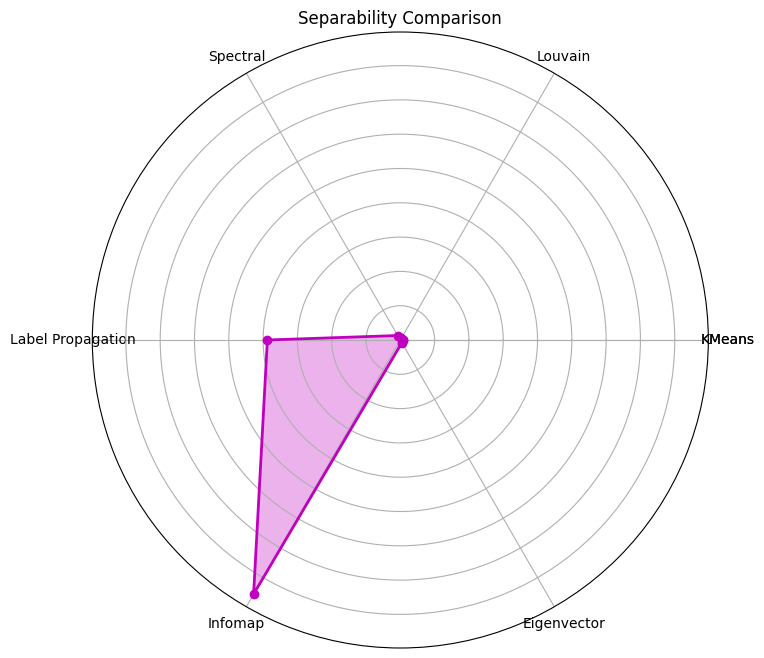}
\caption{Comparison of Separability Score of the algorithms (Higher values indicate well-separated communities)}
\end{figure}

\begin{figure}
\centering
\includegraphics[width=200pt]{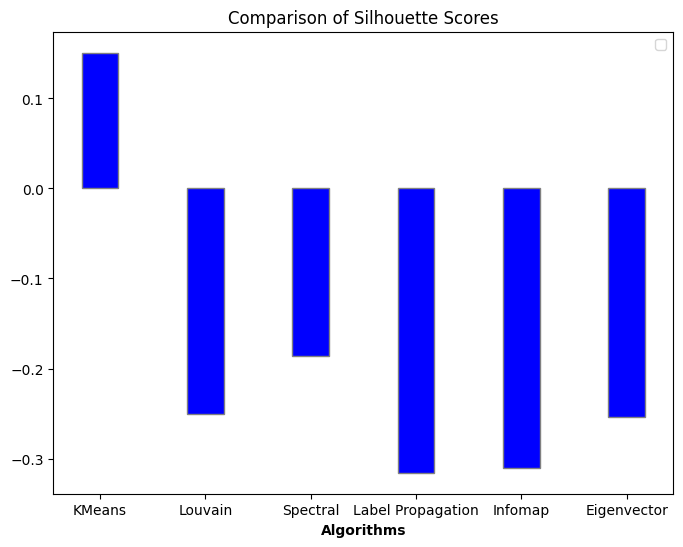}
\caption{Comparison of Silhouette Score of the algorithms (Lower values indicate ambiguous clustering)}
\end{figure}

\section{Conclusion}

In conclusion, our research has rigorously evaluated five prominent community detection algorithms across six key metrics: Modularity, Normalized Cut Ratio, Compactness, Calinski Harabasz Score, Separability, and Silhouette Score. Through this comprehensive analysis, we now posses valuable insights into the strengths and weaknesses of each algorithm.

Our findings reveal that while each algorithm excels in certain aspects, there is no one-size-fits-all solution for community detection. Among the algorithms examined, the Louvain Algorithm and Label Propagation Algorithm emerge as particularly robust choices, demonstrating versatility across multiple metrics and applicability to a wide range of scenarios.

However, it is important to acknowledge that community detection remains a complex and evolving field, and further research is warranted. While specific algorithms show promise in their respective metrics, the quest for optimal community detection solutions continues. Researchers and practitioners should consider the unique requirements of their datasets and objectives when selecting an appropriate algorithm for community detection tasks.

In essence, our study underscores the importance of thoughtful algorithm selection and ongoing exploration in community detection research. By leveraging the strengths of various algorithms and embracing a spirit of continuous improvement, we can advance our understanding of complex network structures and unlock new insights in diverse domains.

Our code is available on github -

https://github.com/YashM246/Comparative\_Analysis\_of\_Community\_De
-tection\_Algorithm

\end{document}